\documentclass[journal]{./IEEEtran/IEEEtran}

\usepackage{color}
\usepackage{url}

\ifCLASSINFOpdf
\else
\usepackage{amsmath, amssymb, amsthm}
\usepackage{graphicx, pstricks, pst-node}
\usepackage{ulem}

\begin{document}

\title{Statistical Classification of Cascading Failures in Power Grids}

\author{Ren\'{e}~Pfitzner,
        Konstantin~Turitsyn
        and~Michael~Chertkov
\thanks{R.P. is a student at the Department
of Physics, University of Jena, Germany as well as at New Mexico Consortium and CNLS, Los Alamos National Laboratory, Los Alamos, NM, USA. E-mail: rene.pfitzner@uni-jena.de.}
\thanks{K.T. is at MIT, Department of Mechanical Engineering and at Los Alamos National Laboratory, Los Alamos, NM, USA.}
\thanks{M.C. is at Los Alamos National Laboratory and at New Mexico Consortium, Los Alamos, NM, USA.}}


\maketitle

\begin{abstract}
We introduce a new microscopic model of the outages in transmission power grids. This model accounts for the automatic response of the grid to load fluctuations that take place on the scale of minutes, when the optimum power flow adjustments and load shedding controls are unavailable. We describe extreme events, initiated by load fluctuations, which cause cascading failures of loads, generators and lines. Our model is quasi-static in the causal, discrete time and sequential resolution of individual failures. The model, in its simplest realization based on the Directed Current description of the power flow problem, is tested on three standard IEEE systems consisting of $30$, $39$ and $118$ buses. Our statistical analysis suggests a straightforward classification of cascading and islanding phases in terms of the ratios between average number of removed loads, generators and links. The analysis also demonstrates sensitivity to variations in line capacities. Future research challenges in modeling and control of cascading outages over real-world power networks are discussed.
\end{abstract}


\IEEEpeerreviewmaketitle

\section{Introduction}
\label{sec:introduction}

The power transmission system is one of the greatest engineering achievements of the past century. The power grid system provides electricity 24 hours a day, seven days a week. However, due to its complexity and spatial extent, it is also vulnerable to failures of various sizes and significance. Extreme events, like the infamous August 2003 blackout \cite{NERC2004a} which left a significant part of North Eastern US without electricity, are rare but their costs to the economy and society are enormous. The significance of this subject has stimulated research in this important area, summarized in a recent review paper by Dobson et al. \cite{DobCarLynNew2007}. Well known cascading models include random network models \cite{Watts2002},  initial power flow based models with gradual load increase, maintenance and random failures \cite{02DCLN,CarLynDobNew2004,NedDobKirCarLyn2006}, phenomenological stochastic models \cite{DobCarNew2003, DobCarNew2004}, hidden-failure embedded power flow models \cite{98TPHT,CheThoDob2005}, and recently power flow based models accounting for adaptive control \cite{Bienstock2010}. In this manuscript, we consider a power flow model of cascading failures, applicable to temporal scales shorter than the time of operator-induced Optimal Power Flow (OPF) control. (OPF, redistributing power generation, shedding load and adjusting frequency, is typically executed on the scale of minutes.)

The demand on the transmission grid grows at a pace significantly exceeding maintenance and system reinforcement. Hence the stress on the grid gradually increases, forcing the system to operate close to its capacity. In a system far from saturation (call it the ''grid of yesterday'') fluctuations in demand rarely led to any significant outage, not to mention a devastating cascade.  On the contrary,  in the system under stress (the ''grid of today or tomorrow'') short-term fluctuations in demand can and unfortunately will lead to significant outages. One realistic scenario of interest is of a grid with sufficient penetration of renewable generation.
Main effect of the renewables, considered as negative loads, consists in generating more frequently the super-critical (leading to outages) fluctuations in demand. This consideration motivates us to analyze outages and cascades generated solely by short-term fluctuations in demand.

Our model and approach extends the research begun in \cite{02DCLN,CarLynDobNew2004,NedDobKirCarLyn2006}, in which the first quasi-static and microscopic (i.e. based on power flows and not on an abstract model of stress redistribution) models of cascades were considered. As in \cite{02DCLN,CarLynDobNew2004,NedDobKirCarLyn2006}, tripping of overloaded lines is a significant part of our power flow dynamics. However, our approach is different in what causes the tripping and how it occurs. Furthermore, in contrast to \cite{02DCLN, CarLynDobNew2004} our analysis concentrates solely on the short-time scale. We do not trip multiple overloaded lines at once, focusing instead on the tripping occurred naturally in response to excessive demand. We do not account for external line outages \cite{CarLynDobNew2004}, effects of sympathetic line tripping \cite{NedDobKirCarLyn2006}, and effects of hidden failures \cite{CheThoDob2005}, which are conjectured to be less significant at the relatively short times discussed in this manuscript.
(Comparative study of outages caused by these different types of failures,  and by the combination thereof, will be the subject of a future publication.)

This study is inspired by the actual operational paradigm guiding the grid dynamics on the scale of minutes or even seconds, when an intelligent manual or semi-automatic control (typically including a human decision in the loop) is unavailable or undesirable. We only account for some standard, automatic and system inherent controls, such as voltage control, automatic line tripping and droop control executed at the generators.
The highlights of our method and results are:
\begin{itemize}
\item We study cascading behavior by solving the power flow equations. Unlike the phenomenological (''disease spread''-like) models, 
our model accounts for \textit{non-local} power-flow based dynamic responses of the power system.

\item We propose a realistic cascade algorithm executed over IEEE test beds, and not over abstract (tree-like or random) structures. This results in a realistic spatio-temporal redistribution of overloads. We point out the importance of this course of action since power systems are not arbitrary but designed intelligently (e.g. capable to withstand any $N-1$ contingency).

\item Our cascade algorithm relies heavily on inhomogeneous line overload and tripping, thus resulting in a dynamical change of the network structure and (in the case of significant damage) leading to the emergence of islands. To study these cascades, we then solve the power flow problem on every island independently. Islanding may also result in an extreme form of load shedding,  i.e. blackout of the entire island if the cumulative load exceeds the generation capacity.

\item The only source of randomness in our model is associated with fluctuations in demand. We show that these demand fluctuations, without other random external influences, are sufficient for causing cascades.

\item The outage growth is quantified in terms of the average number of tripped elements (loads, generators and lines) and their dependence on the magnitude of demand fluctuations. We identify four phases of the overload which differ in how the average characteristics compare. (See Fig.~\ref{fig:phases}.)

\end{itemize}

This manuscript is organized in four Sections. Our cascading algorithm is described in Section \ref{sec:cascading_algorithm}. The algorithm requires solving the power flow equations many times, in a way that mimics the actual dynamics of the grid in a quasi-static causal fashion.  We discuss the application of our cascading algorithm to three IEEE systems, consisting of 30, 39 and 118 buses respectively, in Section \ref{sec:results}. The last Section of the manuscript is reserved for Conclusions and discussing the Path Forward.
\section{Cascading Algorithm}
\label{sec:cascading_algorithm}
\begin{figure}[t]
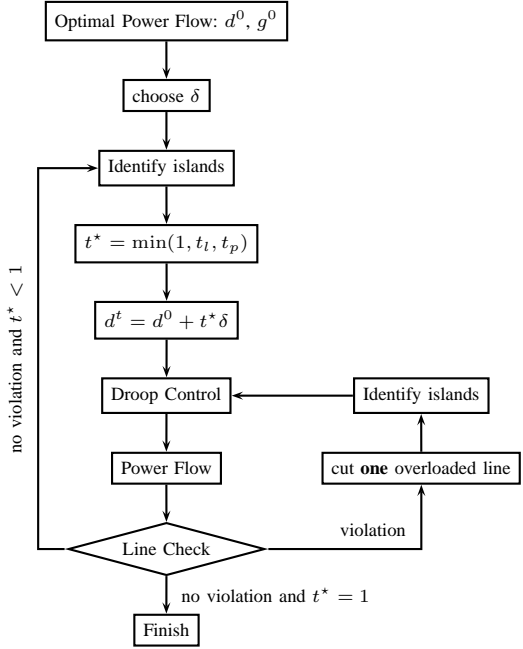

\begin{center}
{\scriptsize
\begin{psmatrix}[colsep=0.05\columnwidth,rowsep=0.5cm]
\rnode{A}{\psframebox{Optimal Power Flow: $d^0$, $g^0$}}\\
\rnode{B}{\psframebox{choose $\delta$}}\\
\rnode{B_1}{\psframebox{Identify islands}}\\
\rnode{C}{\psframebox{$t^{\star}=\min (1,t_l,t_p)$}}\\
\rnode{D}{\psframebox{$d^t=d^0+t^{\star} \delta$}}\\
\rnode{E}{\psframebox{Droop Control}}&\rnode{E2}{\psframebox{Identify islands}}\\
\rnode{F}{\psframebox{Power Flow}}&\rnode{F2}{\psframebox{cut \textbf{one} overloaded line}}\\
\dianode{G}{Line Check}\\
\rnode{H}{\psframebox{Finish}}
\ncline{->}{A}{B}
\ncline{->}{B}{B_1}
\ncline{->}{B_1}{C}
\ncline{->}{C}{D}
\ncline{->}{D}{E}
\ncline{->}{E}{F}
\ncline{->}{F}{G}
\ncline{->}{G}{H}
\naput{no violation and $t^{\star}=1$}
\ncline{->}{F2}{E2}
\ncline{->}{E2}{E}
\ncangle[angleB=-90]{->}{G}{F2}
\naput{violation}
\ncbar[angle=180]{->}{G}{B_1}
\naput[nrot=90]{no violation and $t^{\star}<1$}
\end{psmatrix} }
\caption{Flowchart of the proposed cascading algorithm}
\label{fig:flowchart}
\end{center}
\end{figure}

Our algorithm models cascades triggered by demand fluctuations around the base OPF solution. We consider a quasi-static, sequence of steady states, model.
This algorithm takes an instance of demand for the input and it outputs a solution with balanced load and generation, possibly over a sub-grid of the original grid. A flow chart of the general structure of the algorithm is shown in
Fig.~\ref{fig:flowchart}. The algorithm is rather involved. Thus we find it useful
to begin with a high-level description in the main body, followed
by more detailed expositions of the algorithm sub-tasks in their respective Subsections. (The order of Subsections in this Section is dictated by convenience of the presentation as well as by the interdependence of the material. It does not necessarily reflect the importance of the sub-tasks or their order in execution of the algorithm.)

The algorithm begins by computing the OPF solution $g^0=(g_i|i\in {\cal G}_g)$, using $d^0=(d_i|i\in {\cal G}_d)$ as the base/reference point.
Here, ${\cal G}=({\cal G}_0,{\cal G}_1)$ is the graph of the power grid consisting of a set of vertexes, ${\cal G}_0$, and edges, ${\cal G}_1$. ${\cal G}_g\subset{\cal G}_0$ and ${\cal G}_d\subset{\cal G}_0$ is the subset of nodes with generators and demands, respectively.
Then, the OPF solution is perturbed by a random demand fluctuation, $\delta=(\delta_i|i\in {\cal G}_d)$, drawn from a distribution parameterized by its root-mean-square deviation from the mean, $d^0$. (A detailed discussion of the probability distribution function of the load distribution considered in our simulations can be found in the beginning of Section \ref{subsec:effect_of_random_demand_distributions}.) We apply demand perturbations in a step-wise fashion, with one step corresponding to one cycle of the outer loop in the chart diagram of Fig.~\ref{fig:flowchart}. The number of steps varies with the perturbation and depends on the severity (number of consecutive violations of line or generator constraints which need to be resolved) in the fluctuation of demand. Each step (loop) consists of the following mini-steps: (a) sequential evaluation of the time rescaling parameter $t^{\star}$; (b) redistribution of new loads solely according to droop control (remind that we do not include in this study any control of demand); (c) calculation of Power Flow (PF) solution for the resulting rescaled configuration of demands $d^0+t^{\star} \delta$, laying (in the multi-dimensional space of demands) strictly in between the reference point $d^0$ and the investigated configuration $d^0+\delta$; (d) line check and possible tripping of one violated line based on the DC power flow solution (note that this sub-step involves some randomness in selecting the tripped edge of possibly many violated ones); and (e) checking if the tripping resulted in any islanding. These five mini-steps of the loop are described in detail in Sections \ref{subsec:discrete_time_evolution_of_loads}, \ref{subsec:droop_control}, \ref{subsec:dc_power_flow}, \ref{subsec:line_checking_and_tripping} and \ref{subsec:check_for_islanding} respectively.

It is important to stress at this point that the relatively involved structure of the algorithm mimics the actual microscopic dynamics of the power grid control/adjustment to the demand change.  In particular, the gradual  modification in demand from the reference point to the final contingency, modeled with $t^{\star}$ increasing monotonically from zero to one in a finite number of steps, reflects the physics of the causal response of the grid. Here we assume that the generator-based control of the power flow takes place on the scale of milliseconds, while any change in demand and resulting tripping events occur at much slower pace (measured in seconds, or even longer intervals). Thus the grid has enough time to respond to each individual contingency establishing a quasi-static equilibrium,  modeled in our algorithm by the power flow solver and by the check for islanding steps, after each of the elementary line-tripping or generator-saturation events.
\subsection{DC power flow}
\label{subsec:dc_power_flow}

The power solver takes injection and consumption of powers at all the nodes of the power grid as well as system parameters as input and it outputs voltages and phases at the nodes, as well as the power transmitted over all the links of the grid. Generally, AC power flow is a nonlinear algebraic problem, that becomes a linear problem under a set of additional, so-called Directed Current (DC), assumptions. Our cascading algorithm will work with the most general power solver. However, in this paper we choose to work with a DC solver, which is simpler in implementation and does not require special algorithms for distributing the losses between the generators. The DC solver evaluates
\begin{eqnarray}
&& \forall i\in{\cal G}_0:\quad \sum_{j\sim i}
p_{ij}=\Biggl\{\begin{array}{cc}
 g_i, & i\in{\cal G}_g\\
 -d_i, & i\in{\cal G}_d\\
 0, & i\in{\cal G}_0\setminus({\cal G}_g\cup{\cal G}_d)\end{array}\Biggr.
 \label{eqn:flow_cond}\\
 && \forall \{i,j\}\in{\cal G}_1:\ \ \theta_i-\theta_j=x_{ij}p_{ij}
 \label{eqn:DC_cond}
\end{eqnarray}
where $x=(x_{ij}|\{i,j\}\in{\cal G}_1)$, $g=(g_i|i\in {\cal G}_g)$, $d=(d_i|i\in{\cal G}_d)$, $\theta=(\theta_i|i\in{\cal G}_0)$,
$p=(p_{ij}=-p_{ji}|\{i,j\}\in{\cal G}_1)$ are the vector of line inductances, the vector of powers injected at generators, the vector of demands consumed at loads, the vector of phases and vector of line flows, respectively.
(Here $\{i,j\}$ is our notation for directed edges and $j\sim i$ indicates that $j$ is the graph neighbor of $i$.) Note that to streamline notations, we used an abbreviated version of the DC power flow equations in (\ref{eqn:flow_cond},\ref{eqn:DC_cond}). In particular, we ignore terms associated with tap transformers.
However, n our simulations we utilize the DC PF  solver from the Matlab based MATPOWER package \cite{ZimSanTho2010} taking into account effects of transformers and other devices included in the description of the 30, 39 and 118 nodes IEEE systems.

\subsection{Optimal Power Flow}
\label{subsec:standard_dc_optimal_power_flow}

To set up the system, we solve the standard DC optimal power flow problem finding optimum generator dispatch given the initial load $d^0$, cost functions $f=(f_i|i\in {\cal G}_g)$ for every generator as well as generation power and line capacity constraints. To execute this task we use MATPOWER \cite{ZimSanTho2010}, and cost functions provided in the description of the IEEE systems studied. The DC optimal power flow, in the simplest nomenclature, corresponds to solving
\begin{eqnarray}
\left.\min_{p, g, \theta} \sum_i f_i(g_i)\right|_{
\begin{array}{c}
\mbox{Eqs.~(\ref{eqn:flow_cond},\ref{eqn:DC_cond}), where $d\to d^{0}$}\\
\forall \{i,j\}:\quad |p_{ij}|\leq p_{ij}^{\max}\\
\forall i:\quad g_i^{\min} \leq g_i \leq g_i^{\max}
\end{array}}
\label{OPF}
\end{eqnarray}
for the branch flows, $p$, and generation powers, $g$. The resulting $p^0$, $g^0$ and $\theta^0$ form  the base (reference) solution for our cascading algorithm.
\subsection{Identify islands}
\label{subsec:check_for_islanding}

Our algorithm does not generate a surviving balanced sub-grid at once, but instead it resolves it in steps mimicking dynamics of realistic cascades. The temporal evolution of the surviving sub-grid  is induced by cutting saturated lines, which might also cause the formation of islands, and removing freshly formed but overloaded islands.  We check for islanding (i.e. splitting of the grid into independent components) using the function \textit{grComp} of the MATLAB \textit{grTheory} toolbox  \cite{grTheory}.

If an island is formed, we do all other computations within the cascading algorithm (including DC Power Flow, droop control and calculation of $t^{\star}$) independently for every island.


\subsection{Droop control}
\label{subsec:droop_control}

In the process of evaluating the cascading algorithm, it can happen, due to tripping of overloaded lines, that some loads or generators will become disconnected from the grid or that the grid splits up into islands. Both scenarios require automatic redistribution of generation, done in the so-called droop control fashion~\cite{Kundur1994}.

Droop control is executed at each generator locally in response to an increase or decrease of the system frequency (measured locally as well).
It is required if the grid changes its structure, i.e. following the appearance of new island(s) in the result of line tripping, or if the demand at any node on the island has changed. Here we assume that the power generation, $g_i(+)$, at node $i$ after at least one of these events is
\begin{equation}
g_i(+)=\frac{g_i(-)}{g_\Sigma(-)}d_\Sigma(+),
\label{eqn:droop_control}
\end{equation}
where the newly introduced quantities on the right hand side of Eq.~(\ref{eqn:droop_control}) are the
current power generation, $g_i(-)$, at node $i$;
the total power generation (before droop control), $g_\Sigma(-)=\sum_{j\in \Sigma_g} g_j(-)$, at the freshly formed island, $\Sigma\subset{\cal G}$, the generator belongs to;
and the total power demand, $d_\Sigma(+)=\sum_{j\in\Sigma_d} d_j(+)$, of the island observed after any of the two droop control requiring events. Note that if neither of the two events occurred at the island $\Sigma'$, $d_{\Sigma'}(+)=d_{\Sigma'}(-)$ will hold. Since we always make sure that a stable well-balanced solution demand and generation match, $d_{\Sigma'}(-)=g_{\Sigma'}(-)$ should hold and we arrive at $g_{i}(+)=g_{i}(-)$ in the result.

Droop control is executed at all the generators of the grid simultaneously.
Note that the ratio on the rhs of Eq.~(\ref{eqn:droop_control}) changes in the process of our discrete event simulations in accordance with the modification of islands. If at some point in the process a generator becomes saturated, we do not include it anymore in the droop control mechanism described above, but instead keep its generation level constant (at the maximum generation capacity). As long as demand and total power generation can be matched, the island persists. However, if the total demand of the island exceeds its total generation capacity, we shut down the entire island. Thus, the transition point from phase two to phase three shown in Fig.~\ref{fig:phases} is associated with the emergence of a statistically significant number of islands which were shut down.

\subsection{Discrete Time Evolution of Loads}
\label{subsec:discrete_time_evolution_of_loads}

We do not increase the demand from $d=d^0$ to $d=d^{0}+\delta$ at once, but instead break  the change into a number of (generally not equally spaced) incremental steps.  Each and every next increment is computed separately, as being associated with only a single element modification of the underlying grid (see types (a) and (b) below). To account for the incremental increase of demand, we generate a monotonically increasing sequence of (fraction) times $t^{\star} \in (0,1]$ of load disturbances, each associated with the new configuration of loads, $d^{0}+t^{\star}\delta$. (We recall that if the grid is islanded, we calculate the discrete time sequence separately for each island and choose the minimum time over all islands as $t^{\star}$.) 
As described below, this time sequence is evaluated analytically thus allowing computationally efficient and accurate discrete time implementation in the algorithm. 

We account for constraint breaking events of the following two types:
\begin{enumerate}
\setcounter{enumi}{1}
\item [\textbf{\alph{enumi})}] Exceeding the local maximum generation power constraint at time $t^{\star}=t_g$.
\setcounter{enumi}{2}
\item [\textbf{\alph{enumi})}] Exceeding the line capacity constraint at time $t^{\star}=t_l$.
\end{enumerate}

The time of locally exceeding generation capacity is
\begin{equation}
t_g=\min_i \left(\frac{g_i^{\max}-g_i(-)}{\frac{g_i(-)}{g_{\Sigma}(-)}\delta_{\Sigma}}\right).
\label{t_g}
\end{equation}
Indeed, $t_g$ is the time when $g_i(+,t_g)\geq g_i^{\max}$ holds for exactly one special generator site
$i\in \Sigma$. Then, the expression for the post-event generation at the special site is
\begin{eqnarray}
g_i(+,t)&=&\frac{g_i(-)}{g_{\Sigma}(-)}d_{\Sigma}(+,t)\\
&=&\frac{g_i(-)}{g_{\Sigma}(-)}[d_{\Sigma}(-)+t\delta_{\Sigma}]\\
&=&g_i(-)+t\frac{g_i(-)}{g_{\Sigma}(-)}\delta_{\Sigma}, \label{eqn:droop_control_element}
\end{eqnarray}
where $d_{\Sigma}(+,t)=d_{\Sigma}(-)+t\delta_{\Sigma}$ is the post-event cumulative demand over the island $\Sigma$, and Eq.~(\ref{t_g}) follows from $g_i(+,t_g)=g_i^{\max}$.

Analogously, the time of exceeding a line constraint is
\begin{equation}
t_l=\min_{(ij)} \left( \frac{p_{ij}^{\max}-p_{ij}(-)}{PF({\cal G},\delta, \Delta g)} \right),
\label{t_l}
\end{equation}
which follows from the following consideration. Let $p(\pm)=\text{PF}({\cal G}(\pm),d(\pm),g(\pm))$ denote the vector of power flows, $p$, over the transmission lines of the grid ${\cal G}(\pm)$, obtained by solving Eqs.~(\ref{eqn:flow_cond},\ref{eqn:DC_cond}) with demand $d(\pm)$ and generation $g(\pm)$, where $\pm$ indicates (as before) relevance to the pre- and post- droop control state.
One derives that, $p(+,t)=\text{PF}({\cal G}(-), d(-)+t\delta, g(-)+t\Delta g)$, where $(\Delta g)_i$ is defined according to the droop control rule, Eq. \eqref{eqn:droop_control_element}. Furthermore, since $\text{PF}(...)$ is linear in $d$ and $g$, one finds that
\begin{equation}
p(+,t)=p(-)+t\cdot \text{PF}({\cal G(-)},\delta,\Delta g),
\end{equation}
thus arriving, under condition that $p_{ij}(+,t_l)=p_{ij}^{\max}$ holds for exactly one line, at Eq.~(\ref{t_l}).

\subsection{Line checking and tripping}
\label{subsec:line_checking_and_tripping}

After solving the DC PF equations, we check for lines with violated constraints. The constraints are stated in the two last lines of the conditions for OPF in Eq.~(\ref{OPF}).\footnote{We remind that this manuscript 
deals with relatively fast outages (developing in minutes or even shorter periods),  when 
line capacity can be modeled as a single valued characteristic (emergency rating, C).}
It can easily happen, that one encounters degeneracy, in the sense that the constraints are violated at more than one line. If this is the case, we do not trip all the lines with violated constraints at once, but instead exclude only one of them and then do droop control and DC power flow again. (See the small loop in the flowchart Fig.~\ref{fig:flowchart}.) The order of exclusion is chosen randomly prior to executing the cascading algorithm and it is maintained the same over all iterations and samplings. We stress that this degeneracy can only happen if at a previous step $t_g<t_l$ held, which implies that we change local power generations \textit{instantly}. This course of actions is an approximation and it would of course be more physical to not increase the generation power instantly, but gradually, and to account for singular line tripping events during this gradual increase. However, since dynamic generator data is very often not known, we choose to follow this simple and reasonable scheme.


From the quantitative analysis point of view, we only count a line as tripped when at some point the line flow exceeds its capacity. This means that if an island is overloaded and thus removed from the grid, we do not count lines within the removed island (which now do not carry any current) as tripped. After all, these lines show no need for maintenance. This is different from the way we account  for tripped generators and demands. When an island shuts down (then cumulative load exceeds generation capacity), we count these unserved demands and the  generators set off-line as tripped.

\section{Results}
\label{sec:results}
In this Section we report the results of testing our cascading algorithm, described in details in Section \ref{sec:cascading_algorithm}, on three IEEE systems.

\subsection{Effect of random demand distributions}
\label{subsec:effect_of_random_demand_distributions}

In this Subsection we report tests on standard IEEE systems with 30 and 39 buses respectively. For these two systems all the important system parameters (maximum generation power, $g_i^{\text{max}}$, maximum line capacities, average demand distribution, $d^{(0)}$) are available in the system specifications documented in \cite{ZimSanTho2010}. We select the average load according to, $d_{\cal G}=0.565 g^{\max}_{\cal G}$, while maintaining the same relative distribution of demands between load buses, as specified in \cite{ZimSanTho2010}. (The re-scaling reflects a typical day scenario for the reference point.) Then we set the distribution of generation according to the optimal power flow solution of Eq.~\eqref{OPF}.
Fluctuations in the demands are generated using the half normal distribution allowing only positive fluctuations, in demand $\forall i\in{\cal G}_0$:
\begin{eqnarray}
{\cal P}(\delta_i)=\left\{
\begin{array}{cc}
\frac{\exp(-(\delta_i)^2/(2d_i^0\Delta))}{\sqrt{\pi d_i^0\Delta/2}}, & d_i^0+\delta_i>d_i^0\\
1/2, & d_i^0+\delta_i=d_i^0\\
0, & d_i^0+\delta_i<d_i^0\end{array}\right.
\label{half-normal}
\end{eqnarray}

\begin{figure}[t]
\includegraphics[width=1\columnwidth]{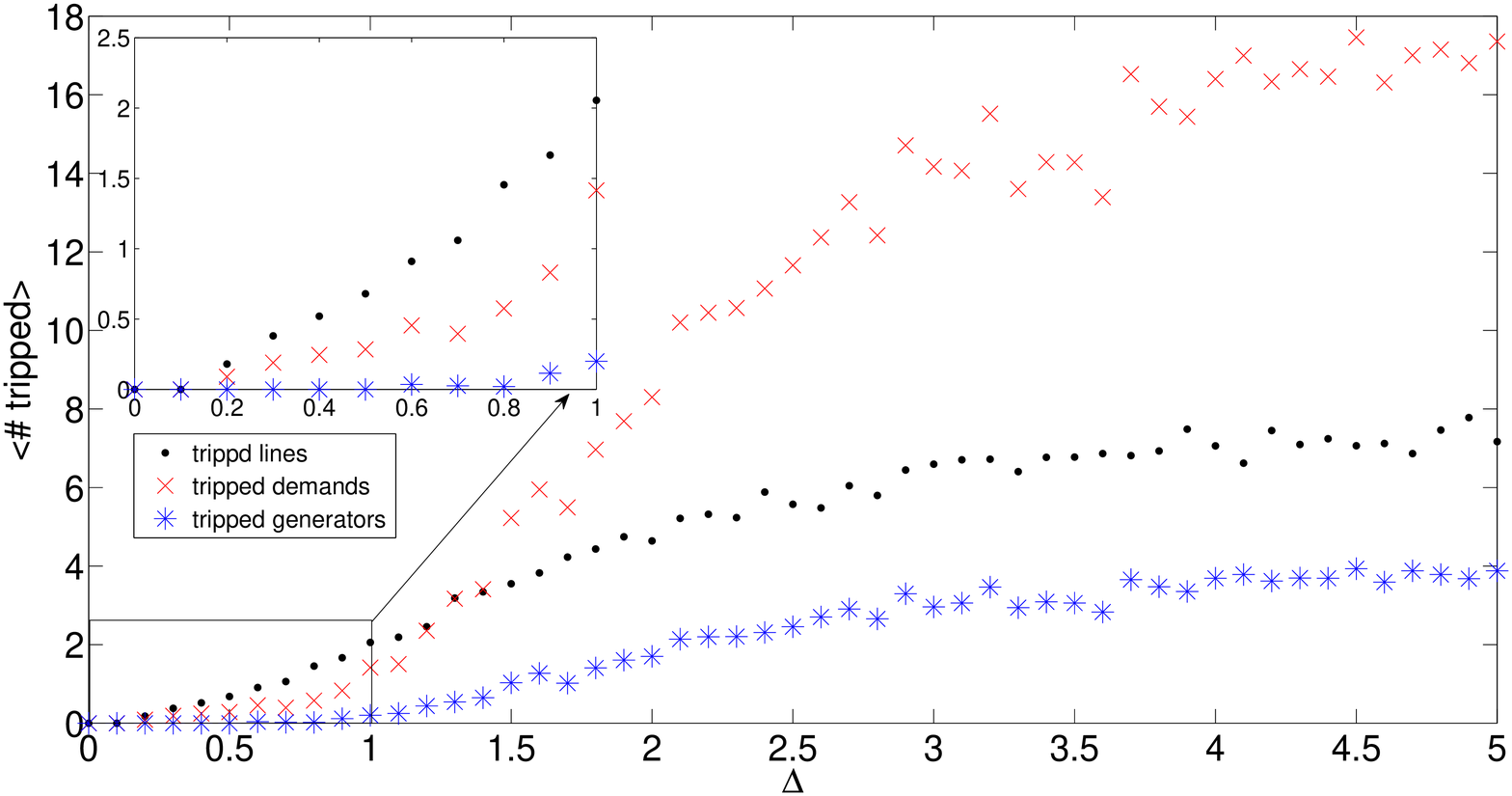}
\caption{Average characteristics of outages in the 30 bus network induced by fluctuations in demand explained in the text. Every data point presents the average over 200 i.i.d. samples using the specified distribution. $d_{\Sigma}^{(0)}=0.565g_{\Sigma}^{(\textit{max})}$. We observe three transition points. Prior to reaching the first transition point the grid is resilient to fluctuations in demand. In between the first and second transition points the probability of having an outage increases slowly. It turns out that the outage in this regime corresponds to tripping of two lines (due to overload) followed by islanding of the adjusted demand node. Here, one observes no cascades yet but only increased probability of line and demand tripping (as also witnessed by the low slope of tripped demands and the stress diagram of the system shown in Fig.~\ref{fig:stress_30bus}). Passing the second transition point indicates emergence of a macroscopically significant number of tripped generators, which also results in a faster rise of demand tripping and signifies the start of cascades. In this system we also note a third transition point at which the number of tripped demands exceeds the number of tripped lines, thus indicating that significant number of the unserved demands belong to islands left without power.}
\label{fig:30bus_dist_a}
\end{figure}
\begin{figure}[t]
\includegraphics[width=1\columnwidth]{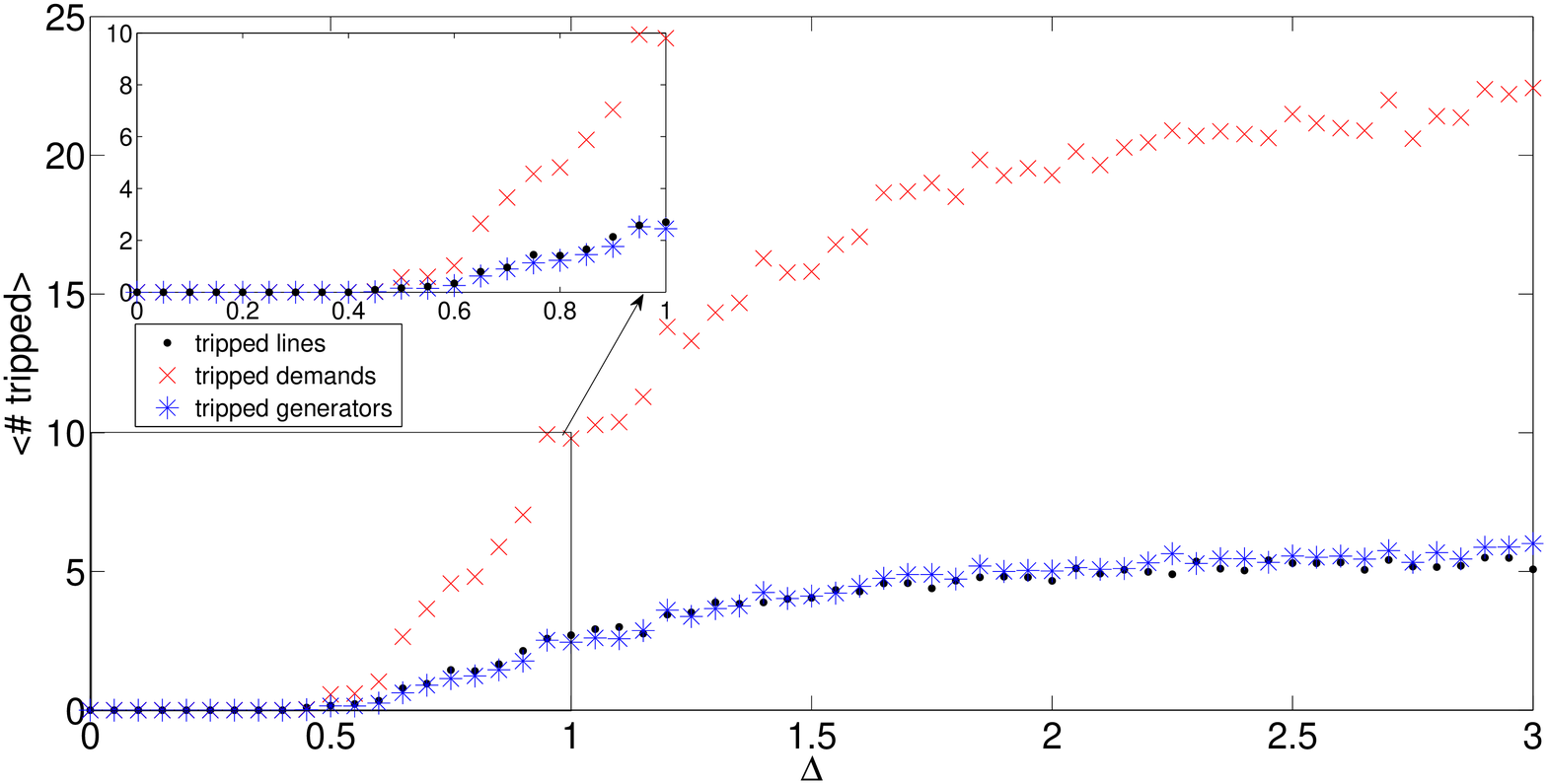}
\caption{Outages in the 39bus network induced by disorder in the demands. Every data point presents the average over 200 samples of Eq.~(\ref{half-normal}). $d_{\Sigma}^{(0)}=0.565g_{\Sigma}^{(\textit{max})}$.}
\label{fig:39bus_dist_a}
\end{figure}

\begin{figure}[t]
\includegraphics[width=1\columnwidth]{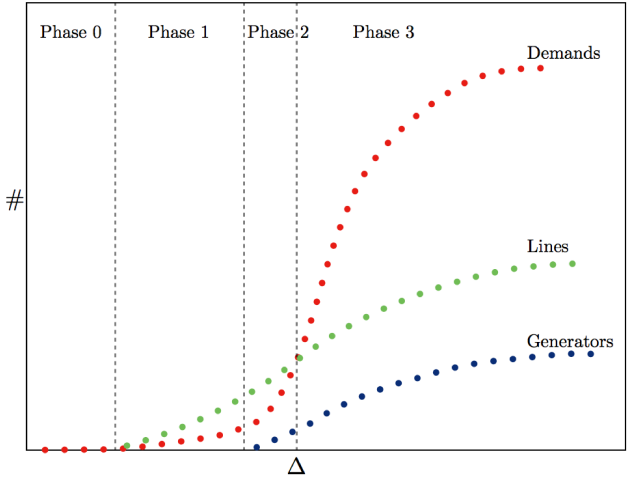}
\caption{Illustration of possible cascade phases. In phase 0 the grid is resilient against fluctuations in demand. Phase 1 shows tripping of demands due to tripping of overloaded lines. This has a overall ''de-stressing'' effect on the grid. In phase 2 generator nodes start to become tripped, mainly due to  islanding of individual generators. With the early tripping of generators the system becomes stressed and cascade evolves much faster (with increase in the level of demand fluctuations) when compared with a relatively modest increase observed in phase 1. Outages in phase 3 are associated with removal from the grid of complex islands, containing both generators and demands.
}
\label{fig:phases}
\end{figure}

Running our algorithm and observing average characteristics of outages, we see the emergence of three distinct cascading phases, illustrated in Figure \ref{fig:phases}. Phase \#0, described by small demand fluctuations,  does not lead to any significant damage. Phase \#1, described by modest fluctuations, results in the removal of some number of lines as well as the removal of a few loads (the formation of islands which do not have any generation and are thus blacked out immediately), while generators remain largely unaffected. This phase can have a de-stressing effect on the grid seen in a significant reduction of damage increase with increasing demand fluctuations. Phase \#2, described by sizable fluctuations, is characterized by the appearance of some tripped generators (surrounded by tripped lines), while the relations, $\langle\#\mbox{ of tripped lines}\rangle
>\langle\#\mbox{of tripped loads}\rangle >\langle\#\mbox{ of tripped generators}\rangle$, remain valid. Finally, phase \#3, described by large demand fluctuations, is characterized by multiple islands, of which a sizable $O(1)$ portion is outaged:
$\langle\#\mbox{ of tripped loads}\rangle
>\langle\#\mbox{of tripped lines}\rangle >\langle\#\mbox{ of tripped generators}\rangle$. Typical instances, contributing to phases \#1-\#3, develop in multiple sequential steps, and as such can all be interpreted as cascades of severity increasing with the numerical index of the phase.

\begin{figure}[t]
\includegraphics[width=0.48\columnwidth]{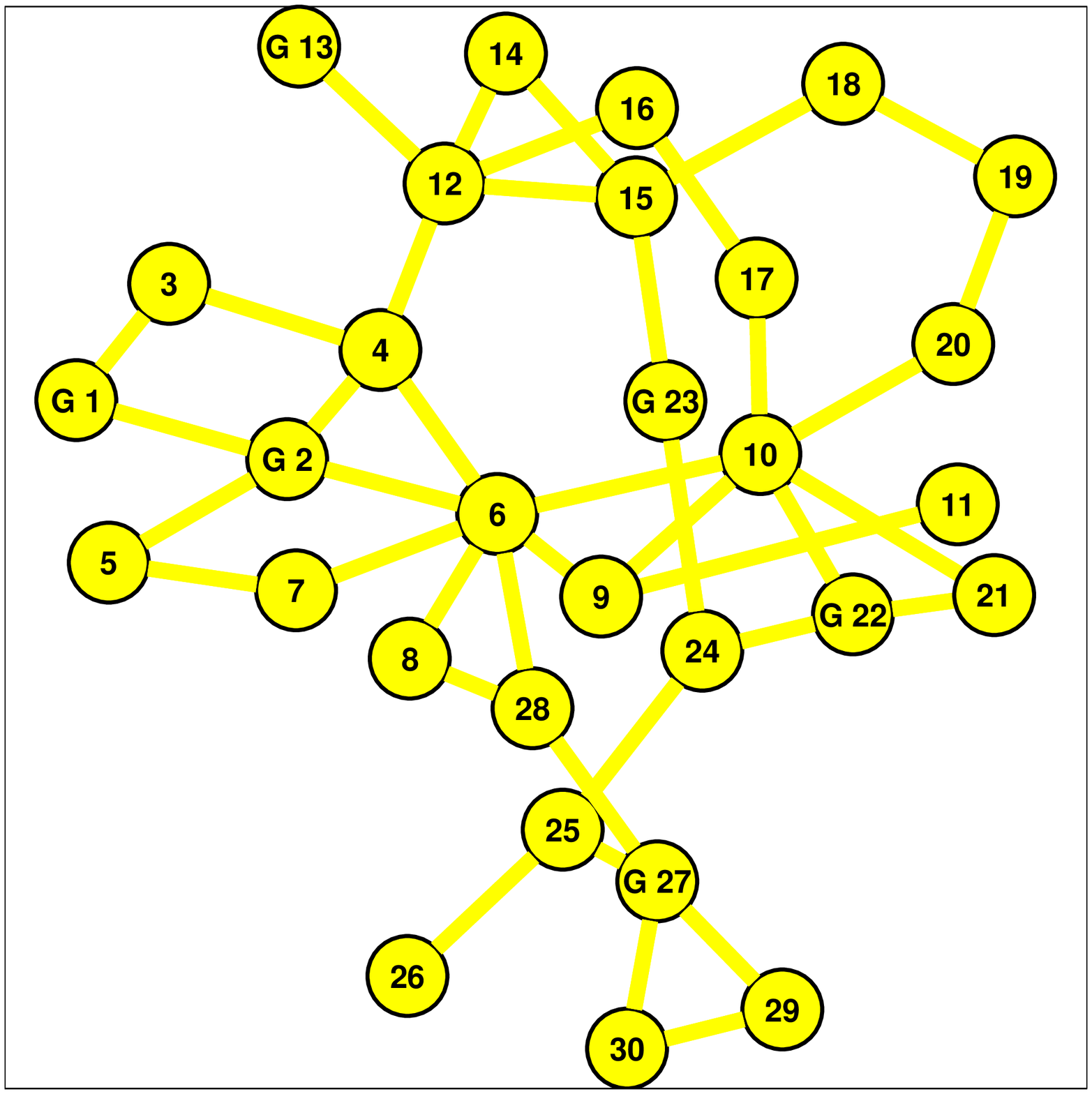}
\includegraphics[width=0.48\columnwidth]{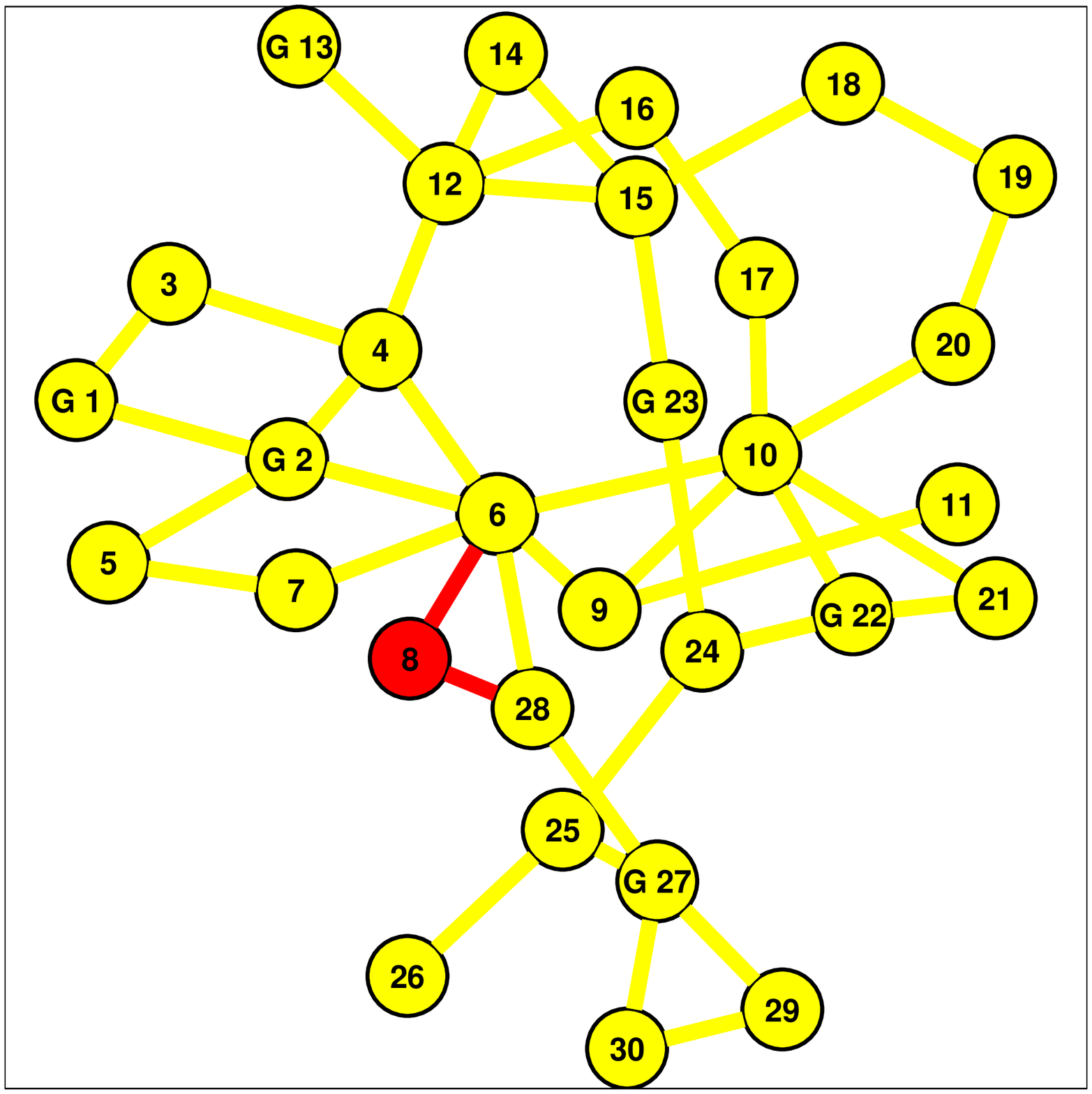}
\includegraphics[width=0.48\columnwidth]{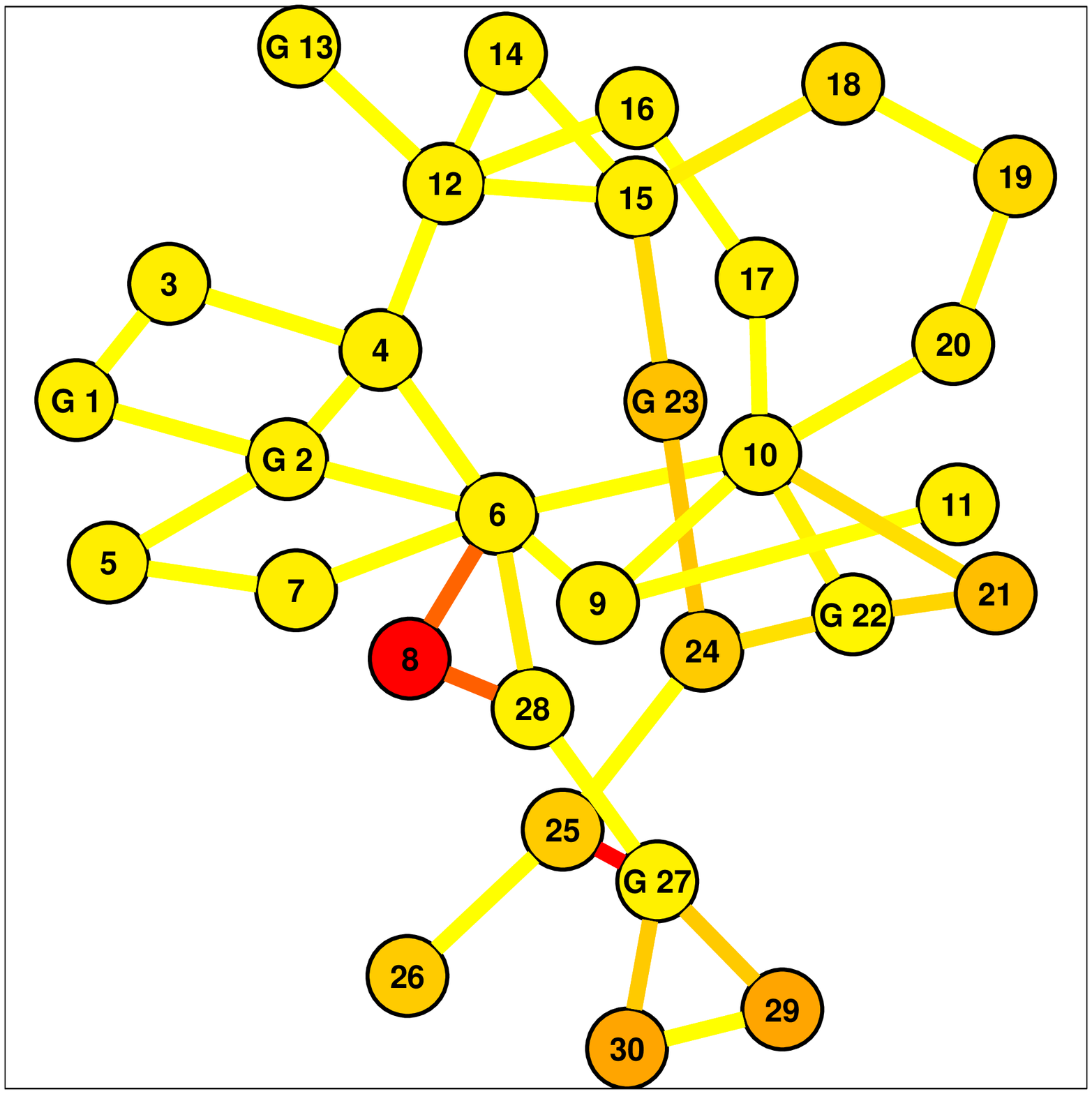}
\includegraphics[width=0.48\columnwidth]{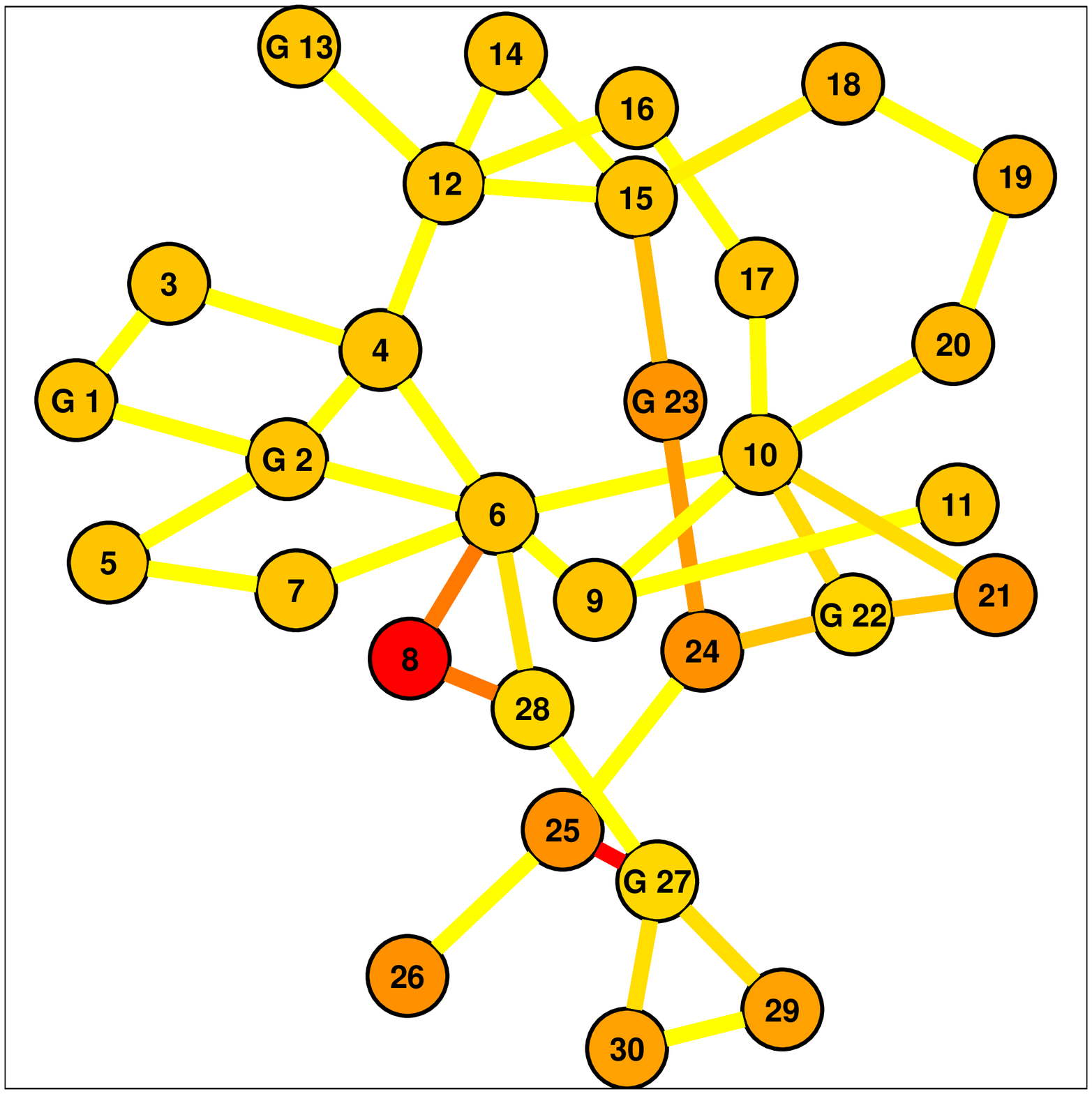}
\includegraphics[width=0.48\columnwidth]{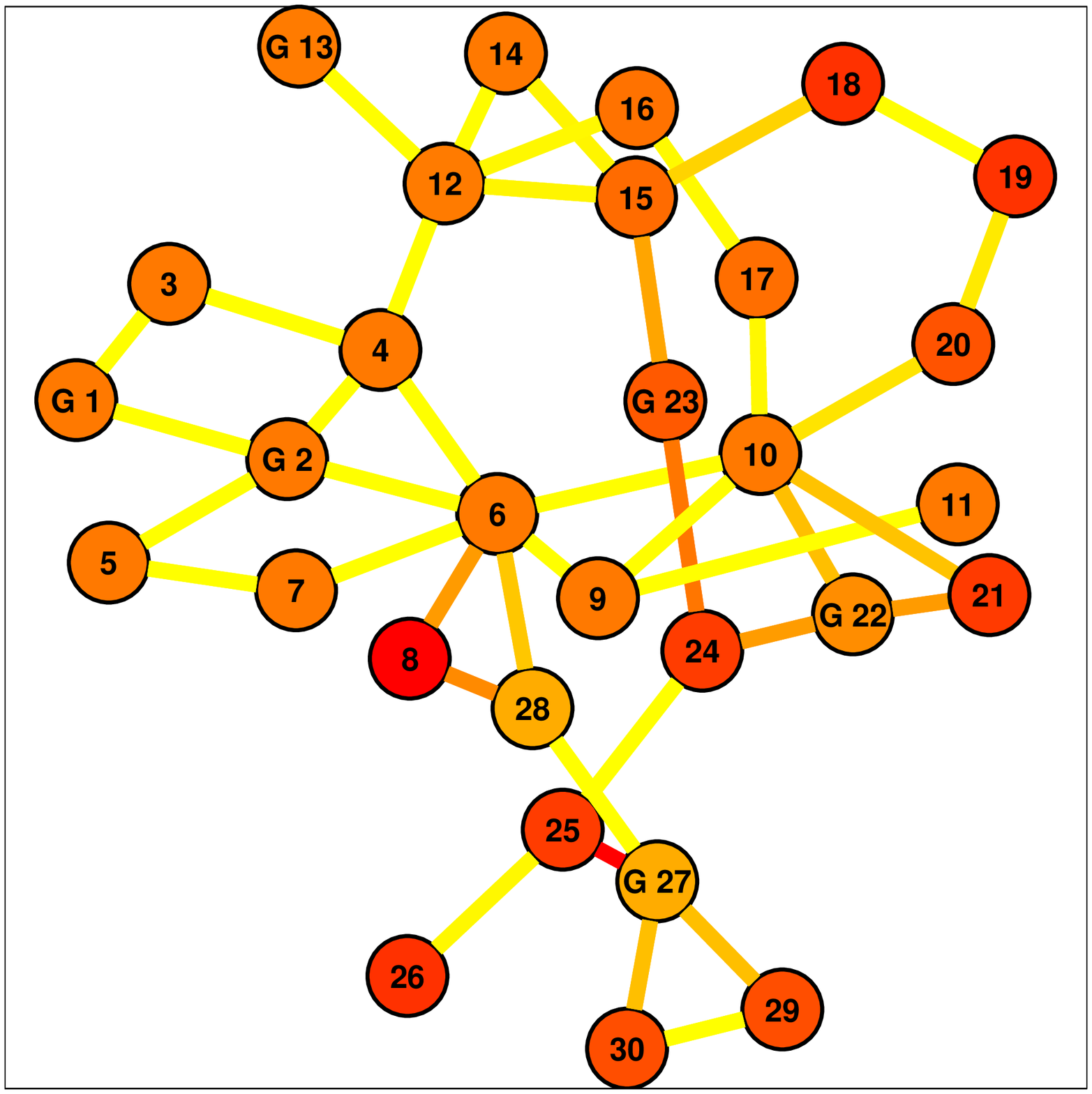}
\includegraphics[width=0.48\columnwidth]{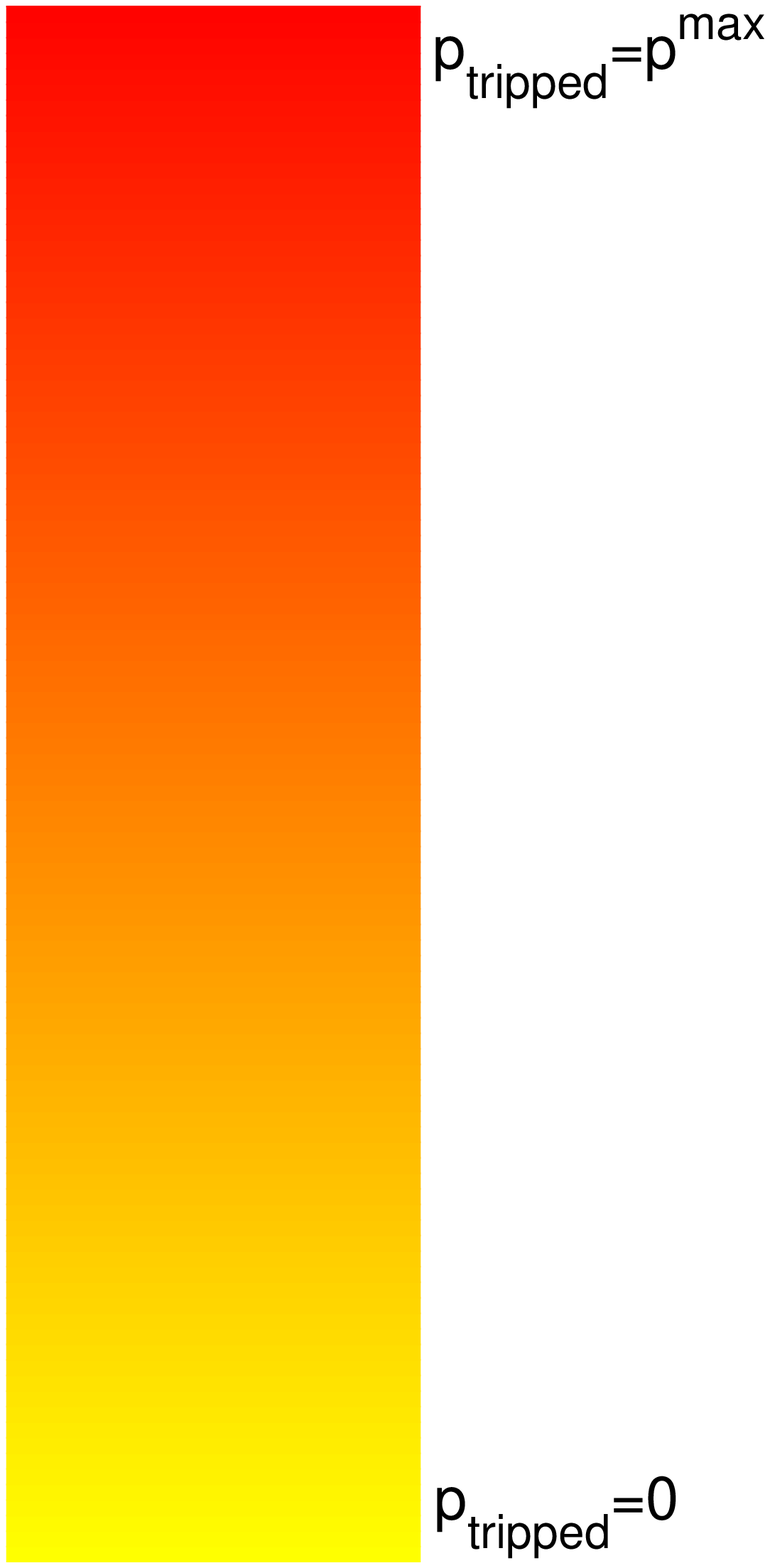}
\caption{Stress diagram of the 30-bus system, corresponding to Fig.~\ref{fig:30bus_dist_a}, using the specified distribution. From top left to bottom right stress in the system is increased: $\Delta=0.1$, $\Delta=0.2$, $\Delta=0.9$, $\Delta=1.2$ and $\Delta=2.0$. Buses labeled $G\#$ are generator buses, all other are load-only buses. The average probability of tripping a line or a bus (over 200 samples) is color-coded for every component. Every instance is normalized by the maximum tripping probability $p^{\text{max}}$ of a component. Yellow (light) means small and red (dark) means maximum tripping probability. (See electronic version of the manuscript for color figures.)}
\label{fig:stress_30bus}
\end{figure}

\begin{figure}[t]
\includegraphics[width=0.48\columnwidth]{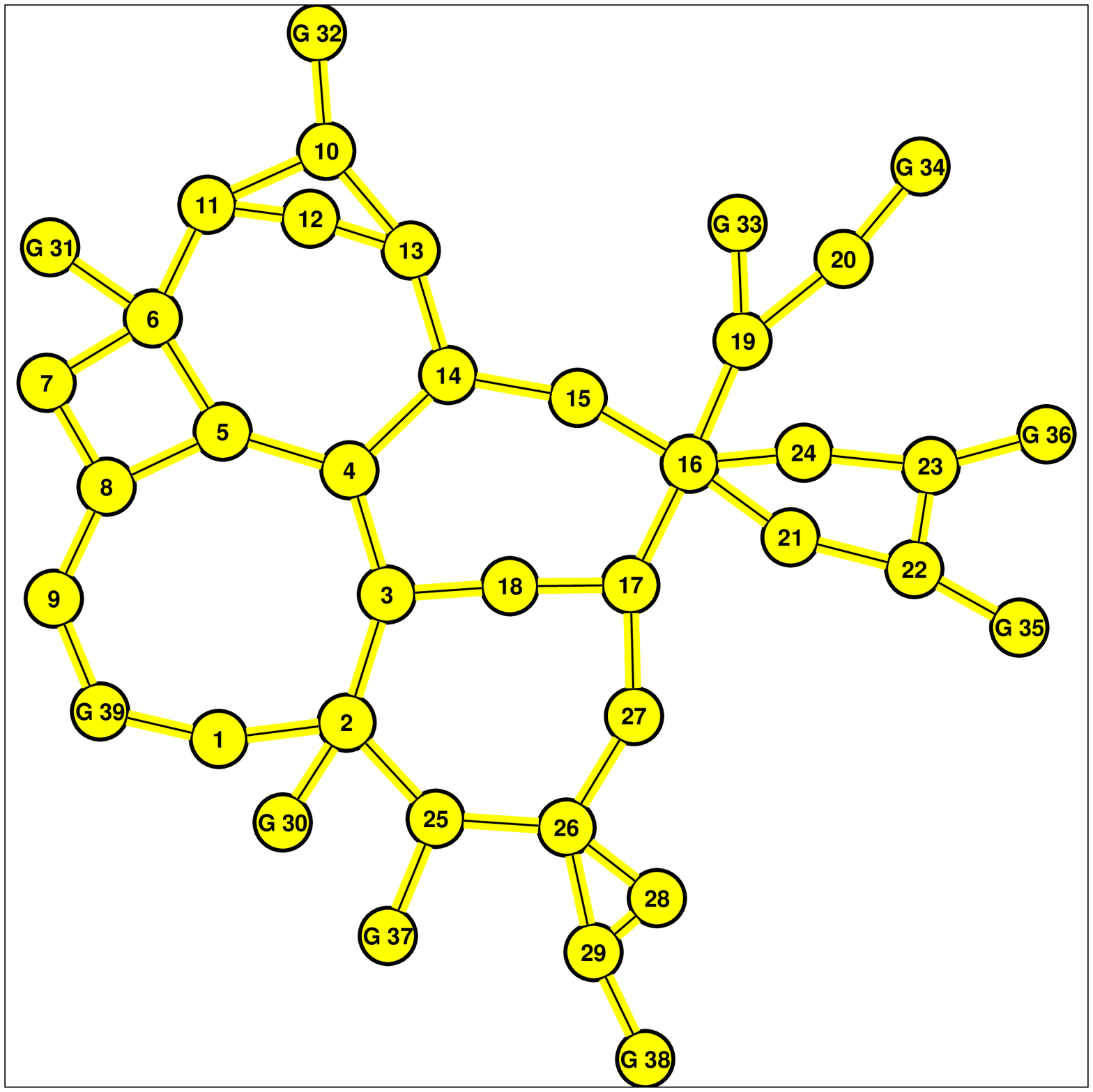}
\includegraphics[width=0.48\columnwidth]{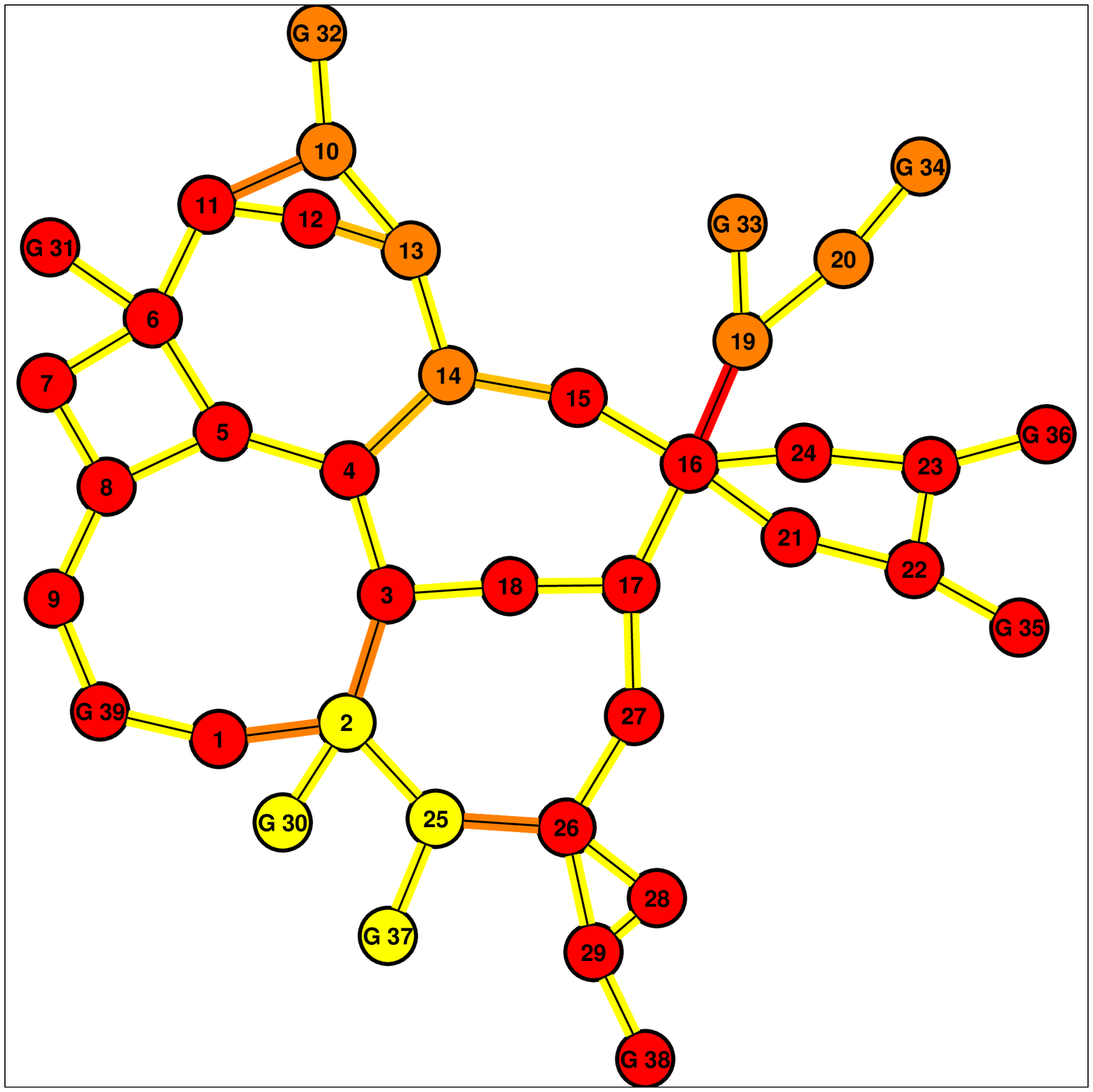}
\includegraphics[width=0.48\columnwidth]{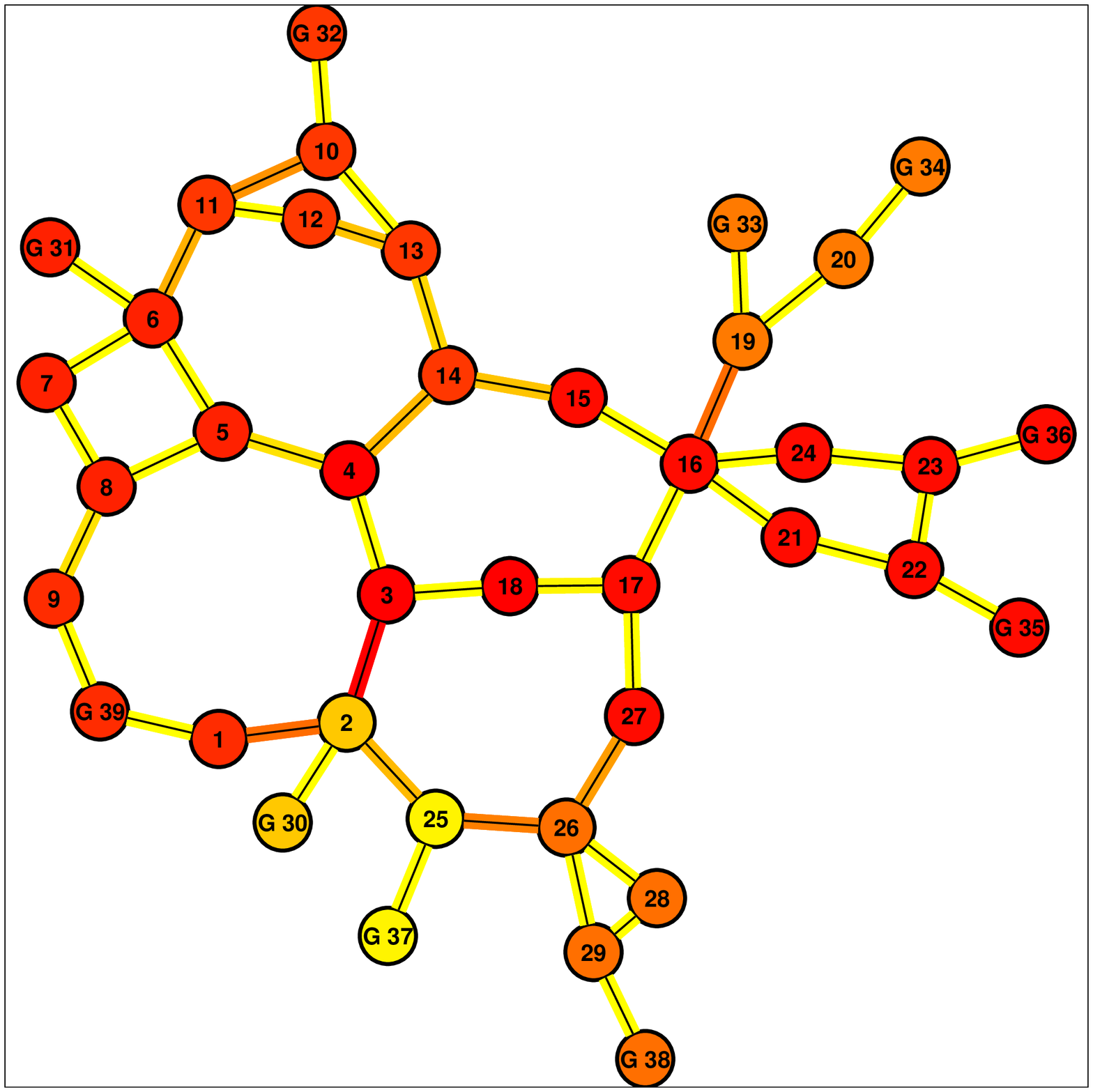}
\includegraphics[width=0.48\columnwidth]{./figs/grid_stress_30_new/node_color_bar.eps}
\caption{Stress diagram of the 39-bus system, corresponding to Fig.~\ref{fig:39bus_dist_a}, using the specified distribution. From top left to bottom right stress in the system is increased: $\Delta=0.3$, $\Delta=0.4$ and $\Delta=0.6$. Buses labeled $G\#$ are generator buses, all other are load-only buses. The average probability of tripping a line or bus (over 200 samples) is color-coded for every component. Every instance is normalized by the maximum tripping probability $p^{\text{max}}$ of a component. Yellow (light) means small and red (dark) means maximum tripping probability. (See electronic version of the manuscript for color figures.)}
\label{fig:stress_39bus}
\end{figure}

Figs. \ref{fig:30bus_dist_a} and \ref{fig:39bus_dist_a} show simulation results on average characteristics of outages, caused
by the distribution of demand Eq. \eqref{half-normal} under different values of dispersion, for the systems of 30 and 39 buses respectively. Although the qualitative forms of the curves observed are quite similar, one also finds interesting differences in the way how cascades evolve in the two systems. Whereas in the 30-bus system the first phase is associated with line tripping also resulting in isolation of a few demands, this state is virtually absent in the 39-bus system, where the cascading behavior begins with the second phase. The increase in the number of unserved demands is here in fact induced by generator tripping. Furthermore phase two and phase three almost coincide in the 39-bus system. However, as the stress-diagrams (see Fig.~\ref{fig:stress_39bus} and Fig.~\ref{fig:stress_30bus}) of both systems show, islanding is an important effect.
The absence of phase one in the 39-bus system appears to be due to the fact that created islands are (a) rather big, (b) include generators, and (c) are stressed and become powerless almost immediately after emergence. In phase one of the 30-bus system, islands are small and immediately removed from the grid (as not containing generators). This early removal of many small islands  has a positive, de-stressing effect on the remaining part of the grid.
The stress diagrams, shown  in Fig.~\ref{fig:stress_39bus} and Fig.~\ref{fig:stress_30bus}, provide additional evidence supporting the aforementioned explanations.

\noindent
A qualitatively similar behavior to the one shown in Figs.~\ref{fig:30bus_dist_a} and~\ref{fig:39bus_dist_a} was also observed for some other choices of demand distributions Eq.~\eqref{half-normal},  e.g. for fluctuations allowing a decrease in demands (yet keeping the total demand positive). Hence one can speculate that the qualitative picture of outage growth induced by increase in load fluctuations is universal.

We would like to comment on one interesting similarity between our microscopic results and the results of Dobson et al. reported for the phenomenological CASCADE model \cite{DobCarNew2003}. The CASCADE model is an abstract representation of the power grid, considering equivalent components failing according to some pre-defined distribution. If a component fails, a certain amount of load is distributed equally to all other components. The model is structureless and as such it carries no explicit relation to the power flow equations. It is argued in \cite{DobCarNew2003} that an increase in the parameter mimicking increase in the total load, results in an abrupt increase in the size of the damage starting at some finite threshold value of the parameter. This observation is akin to the transition from phase one to phase two (see Fig.~\ref{fig:phases}) observed in our microscopic model. Moreover, a similar observation was made recently in the context of yet another phenomenological model discussed in \cite{KadNan2010}. In contrast to the model of \cite{DobCarNew2003},  the one of \cite{KadNan2010} is based on  a network with some spatial structure. Overall,  we conclude that transition from phase one (in which almost all generators are functional)  to phase two  (where a significant, $O(1)$,  fraction of generators is in outage) is observed across the models and hence seems to be an universal feature.

\subsection{Effects of line capacities}
\label{subsec:effects_of_line_capacities}
\begin{figure}[t]
\includegraphics[width=1\columnwidth]{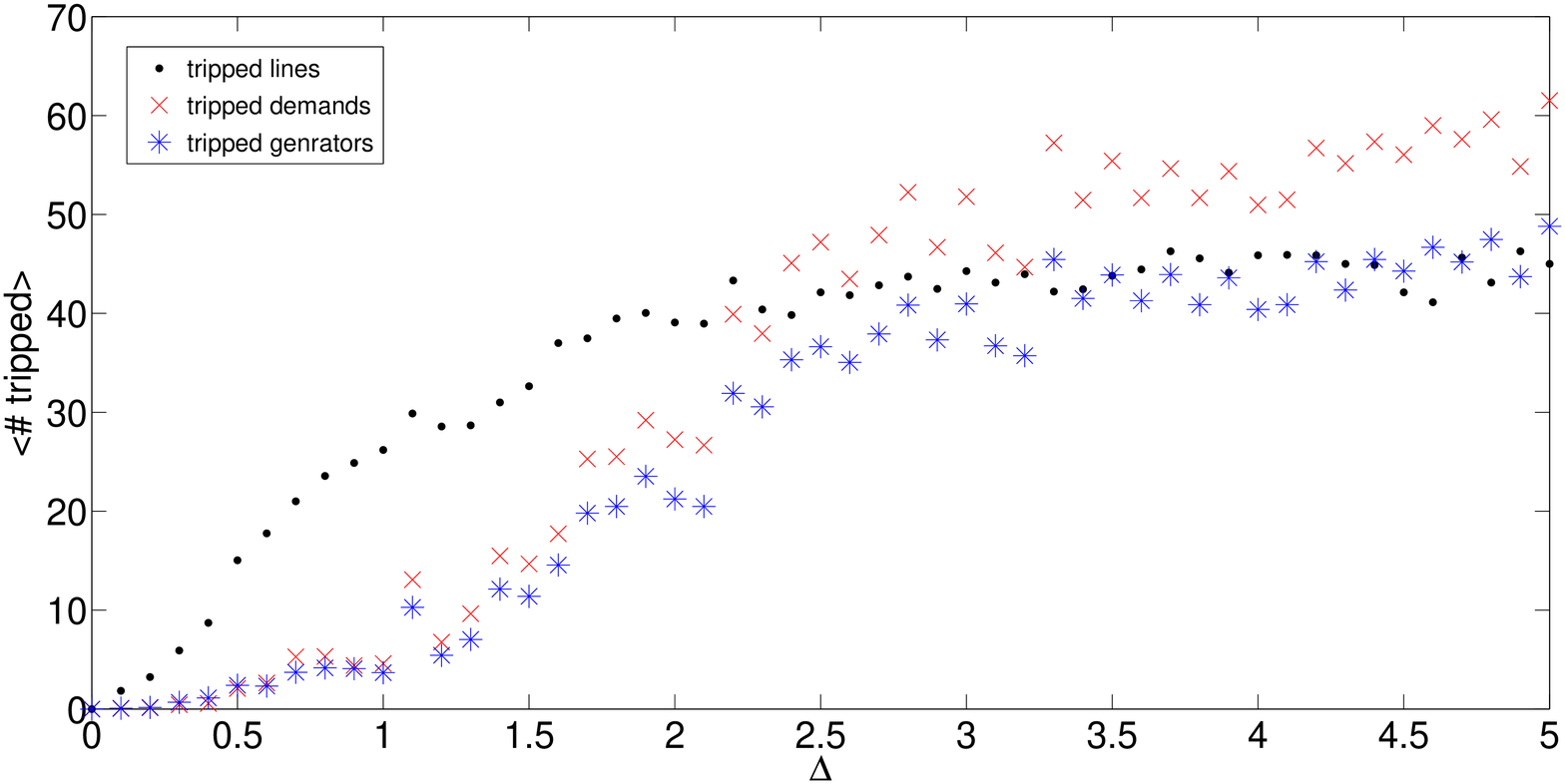}
\caption{Outages in the IEEE 118-bus network induced by disorder in the demands. Every data point presents the average over 25 i.i.d. samples using the specified distribution. $d_{\Sigma}^{(0)}=0.565g_{\Sigma}^{(\textit{max})}$. Since for this systems the IEEE standard does not specify line capacities, we assign line capacities randomly according to the relative-capacity-distribution from the 30-bus case.}
\label{fig:118bus_dist_a_distrib_line_caps}
\end{figure}

In this Section we describe results of our cascading algorithm test on the larger IEEE 118-bus system.
In the contrast with the 30-bus and 39-bus system, specification of the 118-bus system available
in MATPOWER does not have line capacities.
To resolve this problem, we experimented with synthetic distribution of line capacities. We observed that variability in the line capacities affects the dynamics of cascades in a strong way, in particular influencing the structure of emerging islands.

\begin{figure}[t]
\includegraphics[width=0.48\columnwidth]{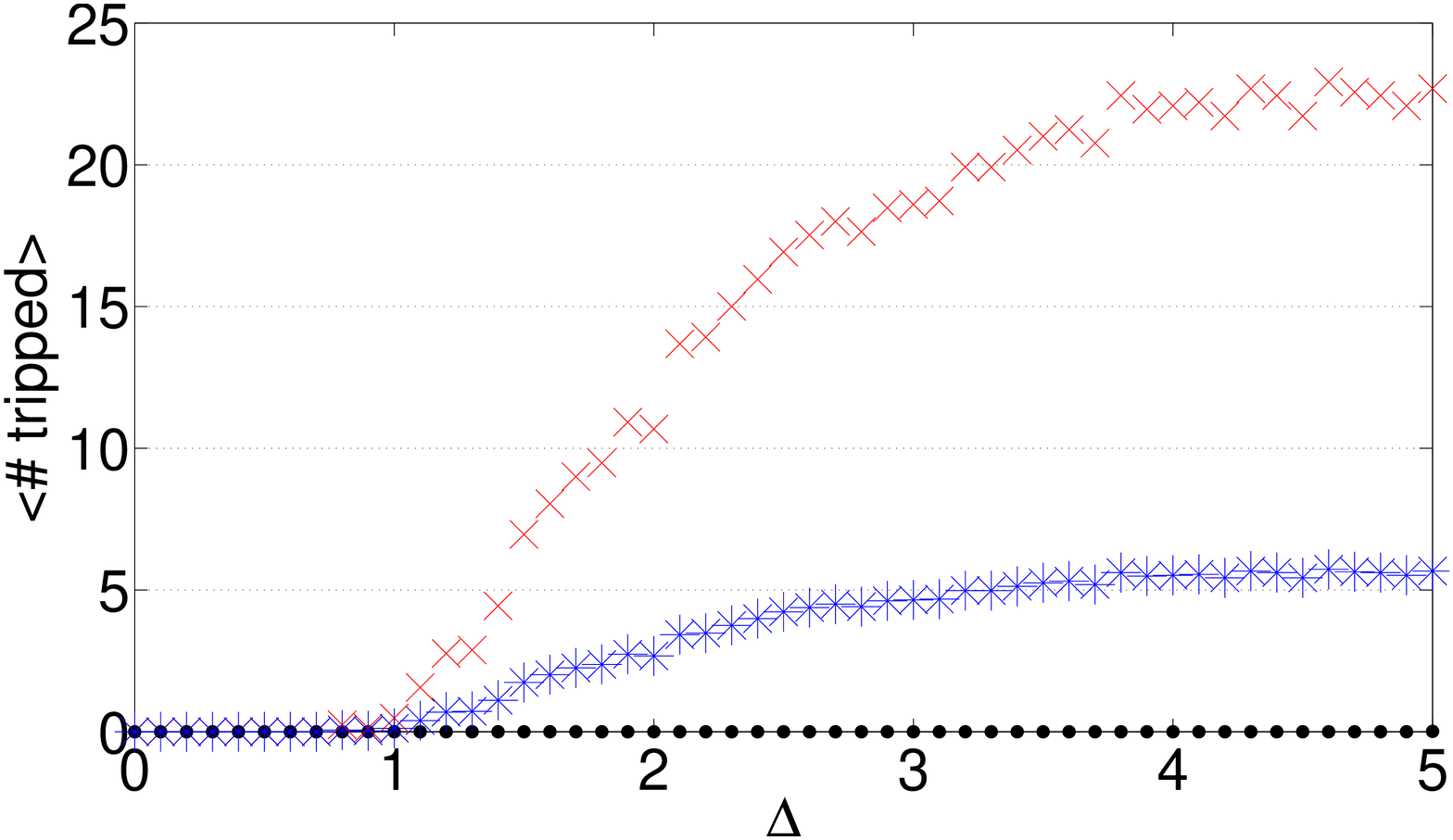}
\includegraphics[width=0.48\columnwidth]{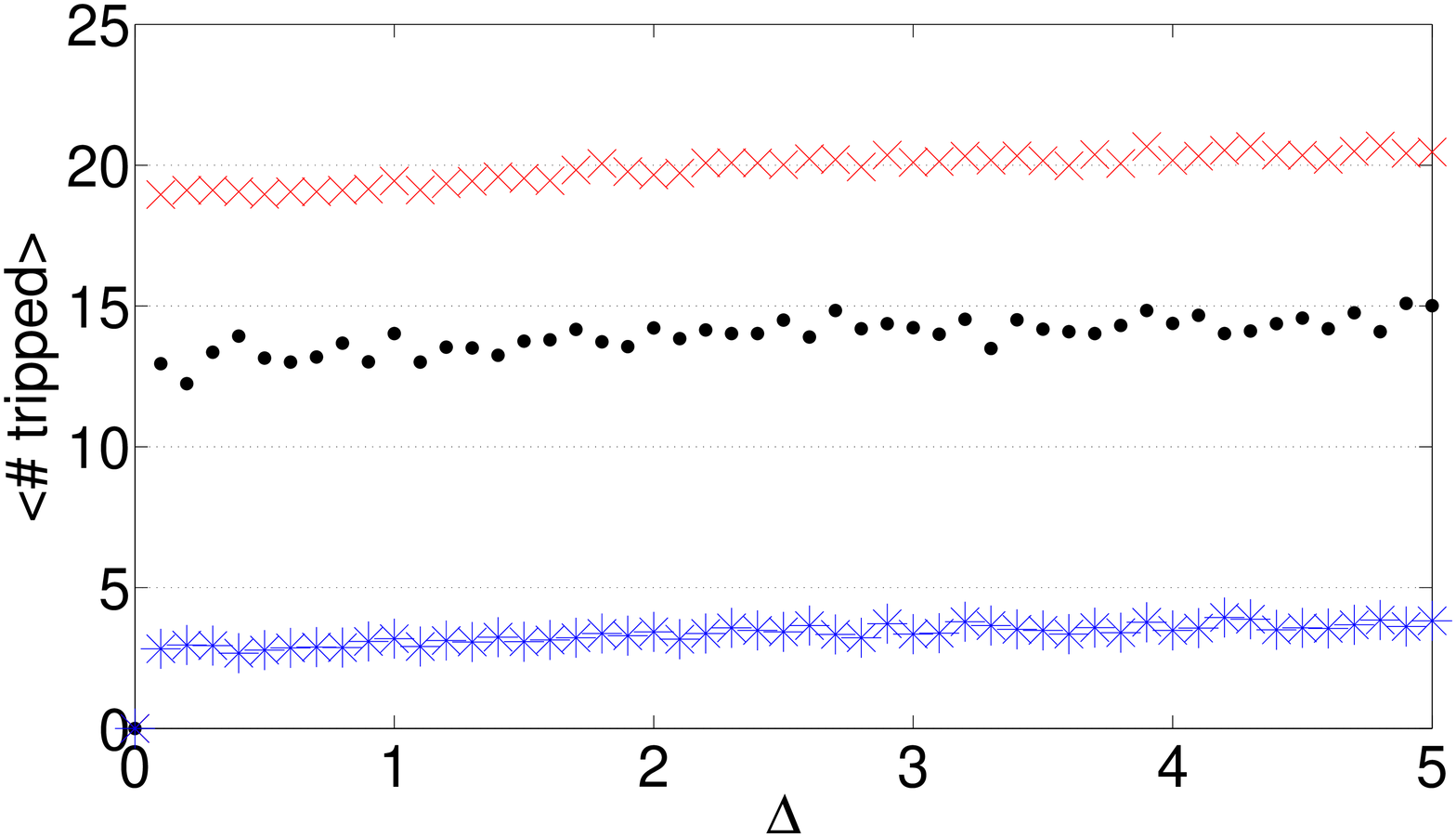}
\includegraphics[width=0.48\columnwidth]{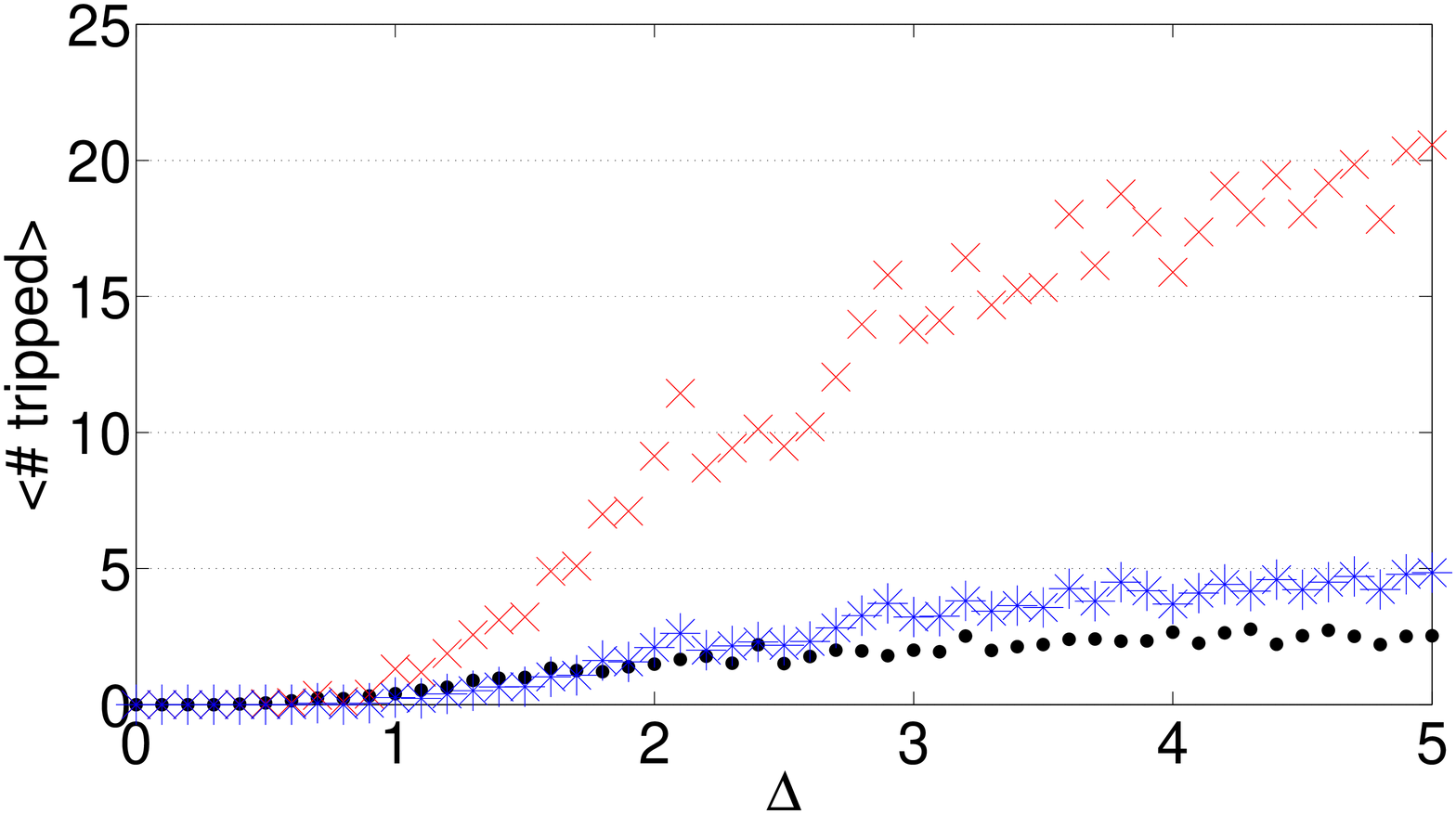}
\includegraphics[width=0.48\columnwidth]{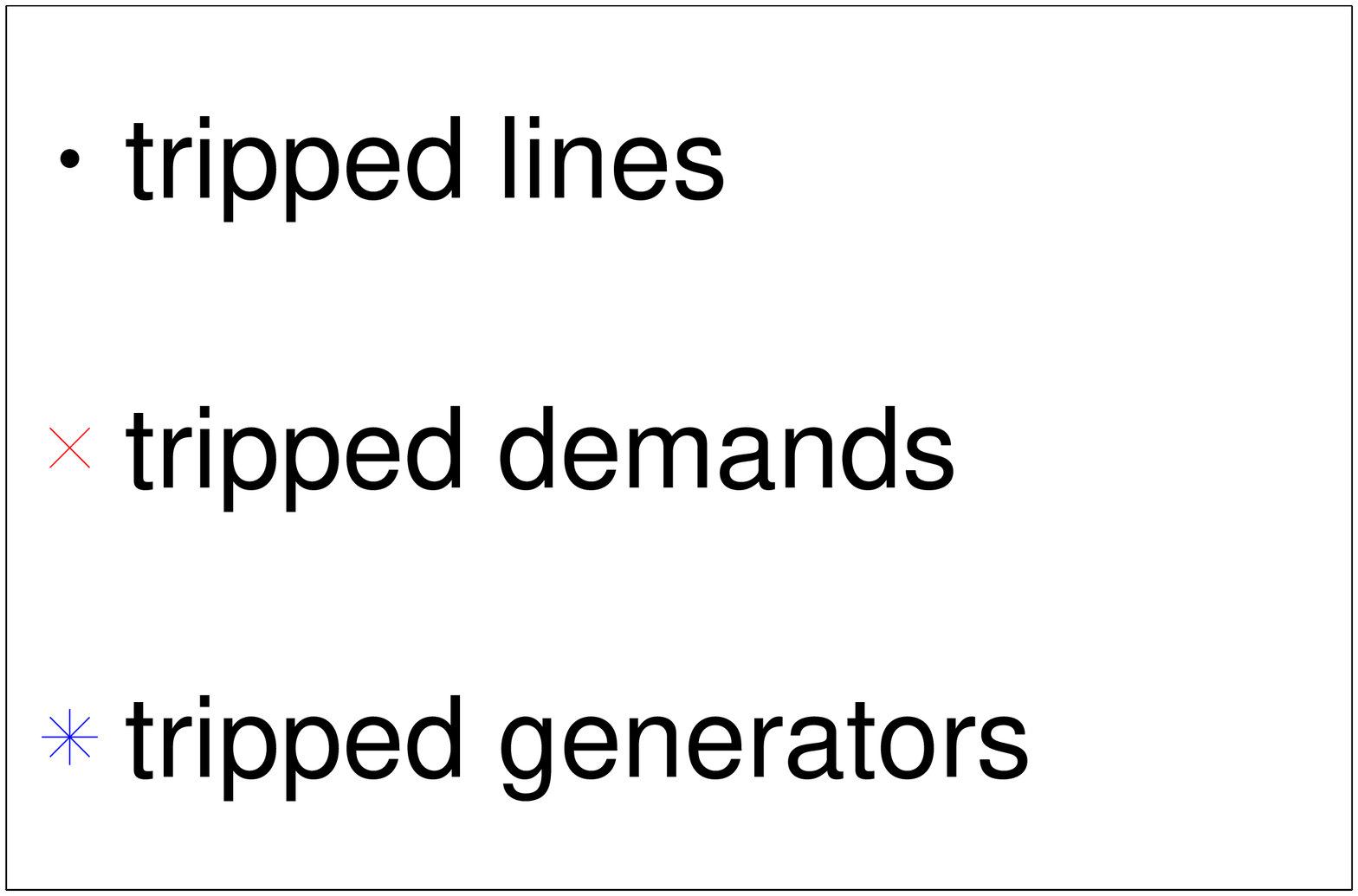}
\caption{Outages in the IEEE-30bus network with univalued line capacities induced by fluctuations in demands. Every data point is a result of averaging over 200 i.i.d. samples from Eq.~(\ref{half-normal}). $d_{\Sigma}^{(0)}=0.565g_{\Sigma}^{(\textit{max})}$. Top left: Set all line capacities equal to the maximum line capacity from the original distribution. Top right: Set all line capacities equal to the minimum line capacity of the original distribution. Bottom left: Set all line capacities equal to the mean of the original distribution.}
\label{fig:line_caps_30bus}
\end{figure}

This strong sensitivity to the line capacity distribution, made first on the 118-bus system,  led us to testing the smaller systems. We show in Figs.~\ref{fig:line_caps_30bus} how the cascading behavior is influenced by different distributions of line capacities in the 30-bus system with 6 generation and 24 demand nodes.
We consider three cases, where all the capacities are set to the same value: equal to the smallest,   largest or  mean characteristics of the original distribution (available in the MATPOWER specification for the system). When the univalued line capacity is maximal (top left figure), no line outage or islanding is observed. Outages in such systems are solely due to generators exceeding their capacities,  which obviously becomes more likely with increase in $\Delta$. Referring to our classification scheme, phase 1 and 2 are absent in this case. This observation is also consistent with results obtained earlier in \cite{DobCarNew2003} or \cite{KadNan2010}.
In contrast, when the univalued line capacity is minimal (top right figure), the cascading behavior is not seen. In this case, any (sensible) initial distribution leads to islanding and blackout of some of these islands. (Note that in this case the whole grid is never outaged completely, instead we reach a stable point with 4 loads and 2 generators remaining for a rather wide variability range in $\Delta$. We attribute this peculiar result to the specific topology of the grid and the initial demand distribution, $d^0$. The islands are formed in a way that two remaining generators provide powers to the remaining loads without exceeding line capacities.)
In the univalued case corresponding to the mean value of the original distribution,
illustrated in the bottom left of Fig.~\ref{fig:line_caps_30bus}, phases 1 and 2 are missing again, suggesting that islanding did not lead to any relief (de-stressing). Looking at all these three univalued examples from a quantitative perspective, we observe that
the average size of the outage at the maximum fluctuations in the demand considered, $\Delta=5$, is significantly larger than in the original case of a realistic distribution of capacities. We associate this negative effect of the univalued capacity with the lack of heterogeneity in the islands formed under stress. To conclude, we find that setting the line capacities to the same value (large, small or averaged) leads to overestimation of the strength of the outage in comparison with the realistic, intelligently designed case.

Therefore, to generate a realistic study of the 118-bus system, we distributed line capacities according to the relative-capacity-distribution of the 30-bus system, i.e. according to
\begin{equation}
\frac{p^{\max}_{ij}}{p^0_{ij}}(\text{30 bus}) \sim \text{Dist}_{ij}(\text{30 bus}).
\label{eqn:cap_dist}
\end{equation}
We obtain the maximum possible power flow (line capacity) over line $\alpha$ in the system as
\begin{equation}
p^{\max}_{ij}(\text{118 bus}) \sim \text{Dist}_{ij}(\text{30 bus}) \cdot p^0_{ij}(\text{118 bus}).
\end{equation}
Of course, our way of distributing capacities in a random fashion according to equation~\eqref{eqn:cap_dist} does not really capture all the features of an intelligently designed system, but serves here as a ''second-order'' approximation (the univalued case being the first order approximation).

Fig.~\ref{fig:118bus_dist_a_distrib_line_caps} shows the resulting outage diagram observed in this synthetic system. These results are consistent with the simulations of the smaller systems and also with the qualitative scheme described in Fig.~\ref{fig:phases}, although phase 1 is more suppressed and phase 2 more extended. Using an intelligent (non-random) assignment of line capacities will lead to a more realistic picture, and should be accounted for un future studies. Note however, that the same simulations conducted with univalued capacities (not shown) give a significantly different picture, qualitatively consistent with the results reported in
Fig.~\ref{fig:line_caps_30bus} for univalued capacity tests in smaller systems.

Based on the results of this Subsection, we conclude that in order to capture realistic cascading effects, like islanding, it is crucial to take the \textit{non-uniformity} of line capacities into account (at least in a ''second-order'' approximation as presented here).

\section{Discussion and conclusion}
\label{sec:discussion_and_conclusion}

In this manuscript, we proposed a new microscopic model of cascades in power grid. The model was tested on  three IEEE systems. We solved the power flow dynamics (in the DC approximation), analyzed structural evolution of the operational part of the grid associated with islanding, and observed the emergence of cascades caused solely by fluctuations in loads. Analyzing the statistics of the damage,  we identified four distinct phases, observed in response to variations in demand fluctuations.
Testing the dependence of the phase structure on line capacities, we observed that selecting line capacities with sufficient variability over the grid is important for capturing realistic dynamics of outages. In particular, our simulations suggest that introducing sufficient variability in line capacities (expressing realm of existing power grids) reinforces the grid, creating multiple islands, and thus making the resulting grid more resistant to a correlated large scale blackout. One observes that a cascade model, which does not account for variations in line capacities, would overestimate the damage.

Our study suggests that to describe  dynamics and statistics of outages in the power grid faithfully, one most account for (a) fluctuations (and eventuall increases) in demand (leading locally and globally to exceeding the generation capacities) and (b) islanding influenced by the distribution of line capacities.

Obviously,  this study constitutes only the beginning of a strategy for analyzing power grid cascades. One natural extension would be to replace the DC power flow solver by a more realistic AC solver. We also intend to study mixed models combining the effects of demand fluctuations with effects of incidental line tripping. Then, with an eye toward aiding efforts in grid reinforcement, we plan to continue our analysis of the effect of capacity inhomogeneities on islanding.  Finally, our long-term goal is to build a novel phenomenological model and theory of cascades based on a detailed microscopic analysis of the type discussed in this manuscript.

\section*{Acknowledgment}
We are thankful to all the participants of the ''Optimization and Control for Smart Grids" LDRD DR project at Los Alamos and the Smart Grid Seminar Series at CNLS/LANL, and especially to S. Backhaus and R. Bent, for multiple fruitful discussions and to D. Bienstock for sharing his working notes \cite{Bienstock2010}. Research at LANL was carried out under the auspices of the National Nuclear Security Administration of the U.S. Department of Energy at Los Alamos National Laboratory under Contract No. DE C52-06NA25396. RP and MC acknowledge partial support of NMC via NSF collaborative grant CCF-0829945. The work of MC on this project was also partially supported by a DTRA basic research grant under Topic 08-D BRCALL08-Per3-D-1-0026.

\ifCLASSOPTIONcaptionsoff
  \newpage
\fi



%

\bibliographystyle{IEEEtran}

\end{document}